\documentclass[a4paper,fleqn,usenatbib]{mnras}
\usepackage[T1]{fontenc}
\usepackage{ae,aecompl}
\usepackage{graphicx}	
\usepackage{amsmath}	
\usepackage{amssymb}	
\usepackage{colordvi}
 \usepackage{ulem}
\usepackage[usenames,dvipsnames]{xcolor}
\usepackage{times}

\newcommand{\gr}{$\gamma$-ray}
\newcommand{\po}{$\pi^0$}
\newcommand{\los}{l.o.s.}

\title[Galactic \gr\ and CR spectra]{Reconciling the diffuse Galactic \gr\ and the cosmic ray spectra}

\author[Nava et al.]
{
Lara Nava$^{1}$
\thanks{E-mail: lara.nava@ts.infn.it}, 
 David Benyamin$^{1}$,
 Tsvi Piran$^{1}$,
 and Nir J.\ Shaviv$^{1,2}$
\\
$^{1}$
Racah Institute of Physics, The Hebrew University of Jerusalem, 91904, Israel\\
$^{2}$
School of Natural Sciences, Institute for Advanced Study, Princeton 08540, NJ, USA
}
\date{Accepted XXX. Received YYY; in original form ZZZ}
\pubyear{2015}

\begin{document}
\label{firstpage}
\pagerange{\pageref{firstpage}--\pageref{lastpage}}
\maketitle

\begin{abstract}
Most of the diffuse Galactic  GeV \gr\ emission is produced via collisions of cosmic ray (CR)  protons with ISM protons. As such the observed spectra of the $\gamma$-rays and the CRs should be strongly linked. 
Recent observations of {\it Fermi}-LAT exhibit a hardening of the \gr\ spectrum at around a hundred  GeV,  between the Sagittarius and Carina tangents, and a further hardening at a few degrees above and below the Galactic plane. However, standard CR propagation  models that assume a time independent source distribution and a location independent diffusion  cannot give rise to a spatially dependent CR (and hence \gr) spectral slopes. Here we consider a dynamic spiral arm model in which the distribution of CR sources is concentrated in the (dynamic) spiral arms, and we study the effects of this model on the $\pi^0$-decay produced $\gamma$-ray spectra. Within this model, near the Galactic arms the observed \gr\  spectral slope is not trivially related to the CR injection spectrum and energy dependence of the diffusion coefficient. We find unique signatures that agree with the {\it Fermi}-LAT observations. This model also provides a physical explanation for the difference between the local CR spectral slope and the CR slope inferred from the average $\gamma$-ray spectrum.

\end{abstract}
\begin{keywords}
diffusion -- cosmic rays -- ISM: general -- Galaxy: general -- gamma rays: ISM
\end{keywords}


\section{Introduction}

Observations of the GeV $\gamma$-ray sky reveal significant information concerning Galactic cosmic rays (CR) at somewhat higher energies. The  $\pi^0$ produced $\gamma$-rays can be used to determine the  CR interaction with the interstellar medium (ISM) at various locations in the Galaxy. The $\gamma$-ray sky can therefore be used as a tool to study CR physics, and in particular, CR acceleration and propagation.  {\it Fermi} observations \citep{Selig2015}, which improved earlier COS B \citep{Rogers1988,Bloemen1989} and EGRET observations \citep{Fatoohi1996}, revealed small but noticeable differences in the $\gamma$-ray spectral index along both different Galactic longitudes and latitudes, particularly so when the lines of sight are tangential to the spiral arms.  We explore here the implications of these observations on models of CR propagation in the Galaxy. 

Spatial variation in the spectra cannot be explained away by variations in the intensity of CRs or variations of the density of the target ISM protons. 
The observations  of a different CR spectrum at different locations imply either that different sources, located at different regions of the Galaxy, have different intrinsic spectra or that CR propagation (in particular the diffusion coefficient) has a different energy dependence at different regions.  
Thus these observations rule out immediately the ``standard" axisymmetric Galactic CR model, or at least imply that it requires the incorporation of new CR physics.
As we show later, the nature of the observed angular dependence is quite complicated (it is asymmetric with respect 
to the Galactic arms) and as such simple modifications of the source or the diffusion inside the arms and outside of them are insufficient to explain the observations. 

With the above in mind, we consider within this context the ``dynamic spiral arm" model that alleviates the assumption of an azimuthally symmetric source distribution and  more realistically assumes that a large fraction of the CR sources are in the vicinity of arms \citep{ShavivPRL,ShavivNewAstronomy,DRAGONspiral,Werner2015}. This model is successful in explaining the variable CR flux as reconstructed from Iron Meteorites \citep{ShavivPRL,ShavivNewAstronomy},  it  naturally explains the so called Pamela anomaly in which the positron to electron ratio increases above 10 GeV \citep{Pamela,DRAGONspiral}, and more recently it also explains the apparent difference in column density inferred from the sub-Iron to Iron and Boron to Carbon measurements \citep{Benyamin2016}.  

One interesting aspect of the spiral arm model is that it requires a smaller halo. This is because its  path length distribution is missing short paths. As a consequence, the same halo size produces less secondaries since the CRs tend to spend less time near the plane, where the density is high. In order to reproduce the observed amount of secondaries the model then requires halo sizes that are typically a few hundred pc high \citep{Benyamin2014,Benyamin2016}. The smaller halo and $^{10}$Be age constraints imply that the diffusion coefficient is also much smaller than the canonical range. Both these effects are supported with recent direct measurements of the CR density in the vicinity of high velocity clouds located outside the Galactic plane, revealing that in some regions the density falls much faster than predicted by standard models, and that it has large horizontal variations \citep{Tibaldo2015}.  It would also explain the indirect inference of the CR density as a function of distance from the Galactic plane using paleoclimate data \citep{Shaviv2014}.

However, without anything else, a {\it static} spiral arm model does not introduce any additional time-scale such that the CR spectral form (but not normalization) should still be location independent. However, \cite{Benyamin2014} considered that the location of the CR sources is progressively moving with the spiral arms. They have shown that advection of the ISM relative to the arm introduces a new effect (and time-scale) at low energies and it naturally explains the rise in the secondary to primary ratio below 1 GeV/nuc. We compare here the predictions of this model with the {\it Fermi} GeV $\gamma$-ray spectral index. One of the interesting points that will be borne out from this comparison is an explanation to what appears as an inconsistency between the local CR spectrum and the spectrum inferred from the average $\gamma$-ray spectral index. This will arise from the effect that the arm dynamics has on the CR spectrum, introducing an upstream/downstream asymmetry around the arms.

We begin by briefly describing  the  all-sky {\it Fermi}/LAT spectral index map \S\ref{sec:obs}, comparing them to standard CR propagation model predictions.  We then  discuss in \S\ref{sec:model}  the dynamic spiral arm CR propagation model we use. We continue in \S\ref{sect:results} with a description of the model results and a comparison with the observations (\S\ref{sec:observations}).  We conclude  with a discussion in \S\ref{sec:discussion}.

\section{Observations}
\label{sec:obs}

Although some recent measurements of the inferred CR spectra in the directions opposite the Galactic centre are consistent with the standard model \citep{Abdo2010}, there are other indications to the contrary. 
COS B measurements of the \gr\ spectra around 300-700 MeV (corresponding to proton energies of typically 7-20 GeV) revealed that the CR spectral index in the direction of the Galactic arms is harder by 0.4$\pm$0.2 than that in other directions \citep{Rogers1988,Bloemen1989}. This was corroborated by EGRET \citep{Fatoohi1996}. Similarly, the electron synchrotron spectrum at 22-30 GHz produced by CR electrons at around 30 GeV is harder in the direction of the arms \citep{Bennett2012}. 

Recently, \citet{Selig2015} analyzed the $6.5\,$yr all-sky data from the {\it Fermi}/LAT in the range of energies between 0.6 and 307.2$\,$GeV. They derived an all-sky spectral index map for the diffuse anisotropic component, giving us the opportunity to test the prediction of the dynamic armed model against observations. 
The map of the \gr\ spectral index $\delta$, estimated in the range $0.85-6.79\,$GeV is shown in Fig.~\ref{fig:comparison_wdata} (lower panel) for latitudes $|b|<7^\circ$.
The spectral index $\delta$ is on average harder in the central region ($-90^\circ<l<40^\circ$), where it also displays a clear variation as a function of the latitude. At small latitudes $|b|<1^\circ$, $\delta$ has an intermediated value $\delta\sim-2.5$, while at larger latitudes a hardening to values $\delta\gtrsim-2.4$ is observed. Larger longitudes are instead characterized by a softer spectrum without a pronounced latitudinal dependence, with an index $\delta\lesssim-2.6$. Thus, the overall spectral slope map appears to have a hard ``oval". 
Two examples of \gr\ spectra (integrated over two different regions of the interest) are shown in Fig.~\ref{fig:comparison_spectra} (dashed lines, from \citealt{Selig2015}).

Although the hard oval can be described as asymmetry between inside and outside the Galaxy, a more natural interpretation given the longitudinal asymmetry is that it is associated with the spiral arms. That is, the spectral slope appears to be harder between the arm tangents (Carina at $\sim -76^\circ$ and Sagittarius at $\sim 50^\circ$). This behaviour is hard to explain in the standard axisymmetric CR diffusion models. If the CR injection spectrum and diffusion coefficient is the same everywhere, there should be no variations in the CR  and  ensuing \gr\ spectra at all, except for the normalization. However, also straight forward extensions changing the characteristic of either the injection spectra or energy dependence of the diffusion coefficient inside the spiral arms cannot explain this behaviour as it would give rise to harder spectra in the direction of the arm tangents, not between them.

\section{The Model}\label{sec:model}
\begin{figure*}
\includegraphics[scale=0.8]{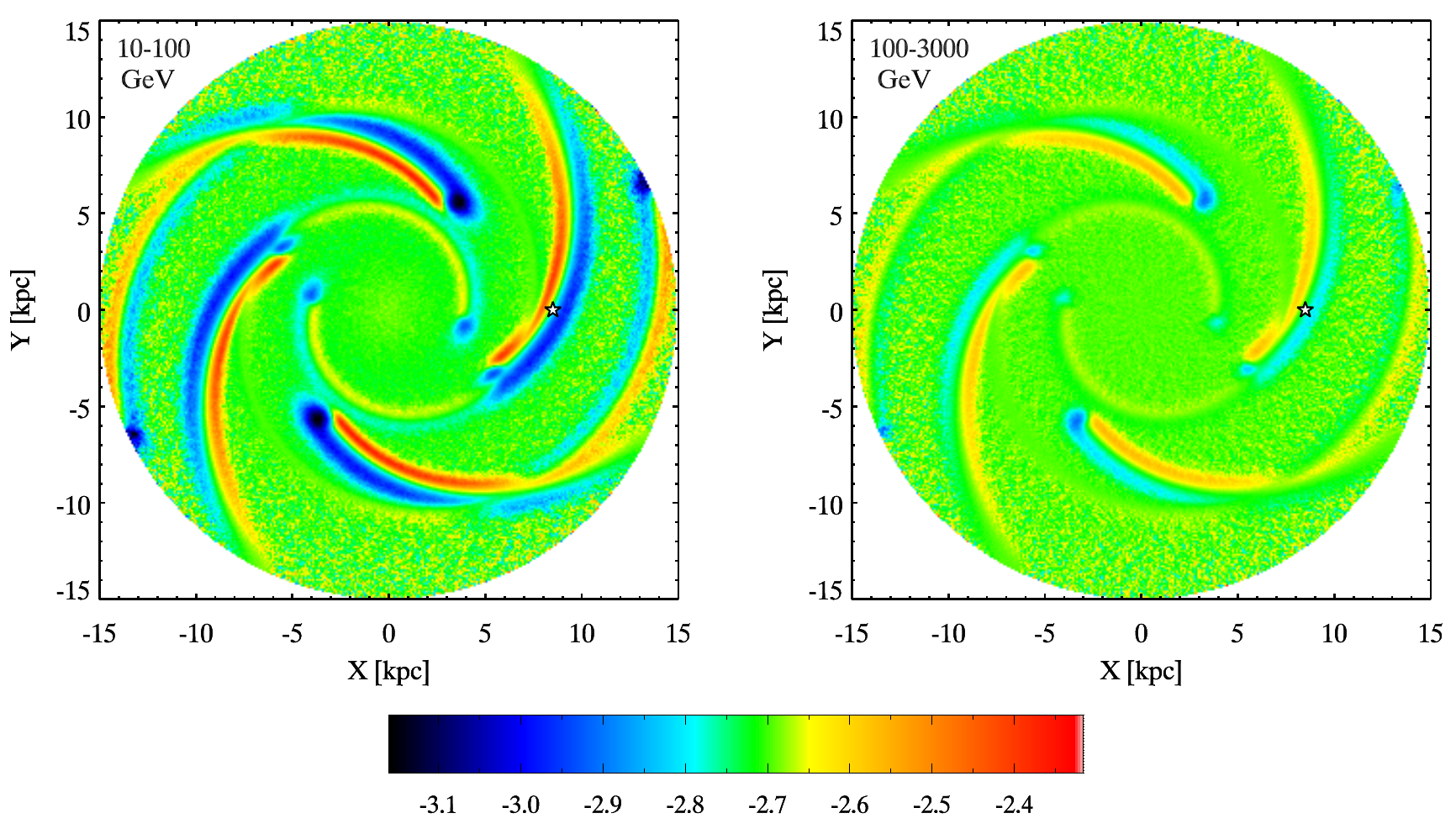}
\caption{A top view map of the Milky Way (at $Z=0$) showing the spectral index of the CR proton spectrum, inferred from simulations, in the range $10-100\,$GeV (left-hand panel) and in the range $100-3000\,$GeV (right-hand panel), as obtained in the small halo model. The Sun is located at $X=8.5\,$kpc and $Y=0$ (star symbol). Note the harder (softer) spectra around the spiral arms that arise from advection of the CR there.}
\label{fig:p_slope_xy}
\end{figure*}

The CR distribution is inferred from the Monte Carlo simulation developed by \cite{Benyamin2014}. While the details of the model can be found in \cite{Benyamin2014}, \cite{ShavivNewAstronomy} and \cite{Pamela}, we summarize here the main features of our model, focusing on those ingredients that are relevant for the study presented in this work.
 CRs of energies up to the knee are assumed to be accelerated at SNR shocks.
Both core collapse and type Ia SNe are considered, with the latter assumed to comprise  20\% of the whole SN population.
We assume here that Type Ia SNe and 10\% of the core collapse SNe are distributed in a homogeneous disc, and take their radial and vertical distributions from \cite{Ferri}.
The rest of core collapse SNe are instead located in the arms---since the progenitors of core collapse SNe live less than $30\,$Myr, they explode not far from their birth sites, which are primarily the spiral arms of the Galaxy \cite[e.g., see][]{Lacey}. 
For the spiral structure we follow the geometry of a superposition of a four-armed set and a two-armed set, with the rational and details described in \cite{Benyamin2014}. This is based on cumulative evidence borne from different empirical results \citep{AL1997,Dame2001,ShavivNewAstronomy,Naoz2007}. 

The CR spatial distribution from the axisymmetric components does not depend on the rotation of the underlying interstellar medium. 
The spiral arms break the axial symmetry.
Furthermore, when considering the propagation from the spiral arms, the rotation of the arms and the rotation of the ISM should be considered. 
This introduces an additional propagation effect, namely,  ``advection" of the interstellar medium relative to the spiral arms. At energies of up to a few GeV/nuc.\ advection can compete with diffusion and a break in the CR spectra is therefore expected in this model. It also naturally explains the apparent rise at low energies in the Boron to Carbon ratio \citep{Benyamin2014}.
The interplay between the advection and diffusion also breaks the symmetry around the spiral arms. 
While diffusion operates in either direction both ahead and beyond the spiral arm, advection is only downstream.
The break in the CR spectrum will manifest itself as a break in the photon spectrum produced in interactions between CRs and the ISM.
The source distribution is then inhomogeneous, non-axisymmetric and time dependent.

We carry out  Monte Carlo simulations  following the full propagation of a large number of CR ``bundles". Their initial location in the Galaxy is randomly chosen according to the source distribution described above. We follow the bundles from their formation until they leave the Galaxy, taking into account energy losses (mainly in the form of Coulomb and ionization cooling).
The diffusion coefficient as a function of rigidity $R$ is described by a single power-law, $D = D_0 (R/R_0)^{s}$, with $s=0.4$, $R_0 = 3$~GeV/nuc, $D_0 = 1.2 \times 10^{27}$cm$^2$/s and a halo size $Z_h=250\,$pc.
We assume that the injection proton spectrum as a function of the momentum $p$ is described by a power-law $dN_{\rm inj}/dp\propto p^{\alpha}$.
For the index $\alpha$, we consider two different values. First, we investigate the standard hypothesis $\alpha=-2.3$, leading to an average CR spectral slope (after diffusion) equal to $\alpha-s=-2.7$. With these values, the Boron to Carbon and Beryllium isotopes ratios \citep{Benyamin2014} and sub-Iron to Iron ratios \citep{Benyamin2016} are recovered.
Then, due to motivations that will be clear later, we will consider the case of a slightly harder CR injection spectrum, with $\alpha=-2.25$.
We also consider a second model, with dynamic arms, but much larger diffusion coefficient  ($D_0=10^{28}$cm$^2$/s) and a halo size of $Z_h=2\,$kpc, which better resembles ``standard" CR propagation models.  

The final CR distribution is recorded on a grid that spans a volume of 30 kpc $\times$ 30 kpc $\times$ 2$Z_h$. It is composed of (0.1 kpc)$^2$ $\times$ (0.01 kpc) sized cells (0.1 kpc in the $X$ and $Y$ directions and 0.01 kpc in the $Z$ direction). In each cell, the spectrum at 43 different energy bins is calculated, from 0.5$\,$GeV up to 500$\,$TeV.
To estimate the local ISM density, we consider all the different components of the gas (neutral, ionized and molecular) and describe their density distribution in the Galaxy according to the parametrization given in \cite{Ferri}.
The local \po\ production and the resulting \gr\ emission are then computed in each cell, from the local CR spectrum and ISM density, following \cite{kamae06}.
Once the \gr\ spectrum is computed in each cell of the simulation box, we can investigate, in our dynamic spiral arm model, how the spectral slope is predicted to change across a spiral arm.
Finally, in order to compare simulations with observations, we integrate the \gr\ emission along different lines of sight (\los) and derive the sky map of the spectral photon index.

\section{Results }\label{sect:results}
Before discussing the results of our simulations, we begin with a qualitative description of the expected features. 
Fig.~\ref{fig:p_slope_xy} shows the CR spectral index spatial distribution as predicted by our simulations, performed with $\alpha=-2.3$. 
The Sun is located at $(X,Y,Z)=(8.5\,\rm kpc,0,0)$, as indicated by the open star symbol.
As expected, the spectrum is harder in the upstream side of the arms, and softer in the downstream side. This is because lower energies CRs are more strongly affected by the advection of the ISM through the arms. The variation of the spectral index across each arm is more evident in the lower energy range of 10-100$\,$GeV (left-hand panel) and becomes less evident at higher energies (right-hand panel), where the relative effect of the advection, as compared to diffusion, is less important. 

\begin{figure*}
\hskip -0.14truecm
\includegraphics[scale=0.8]{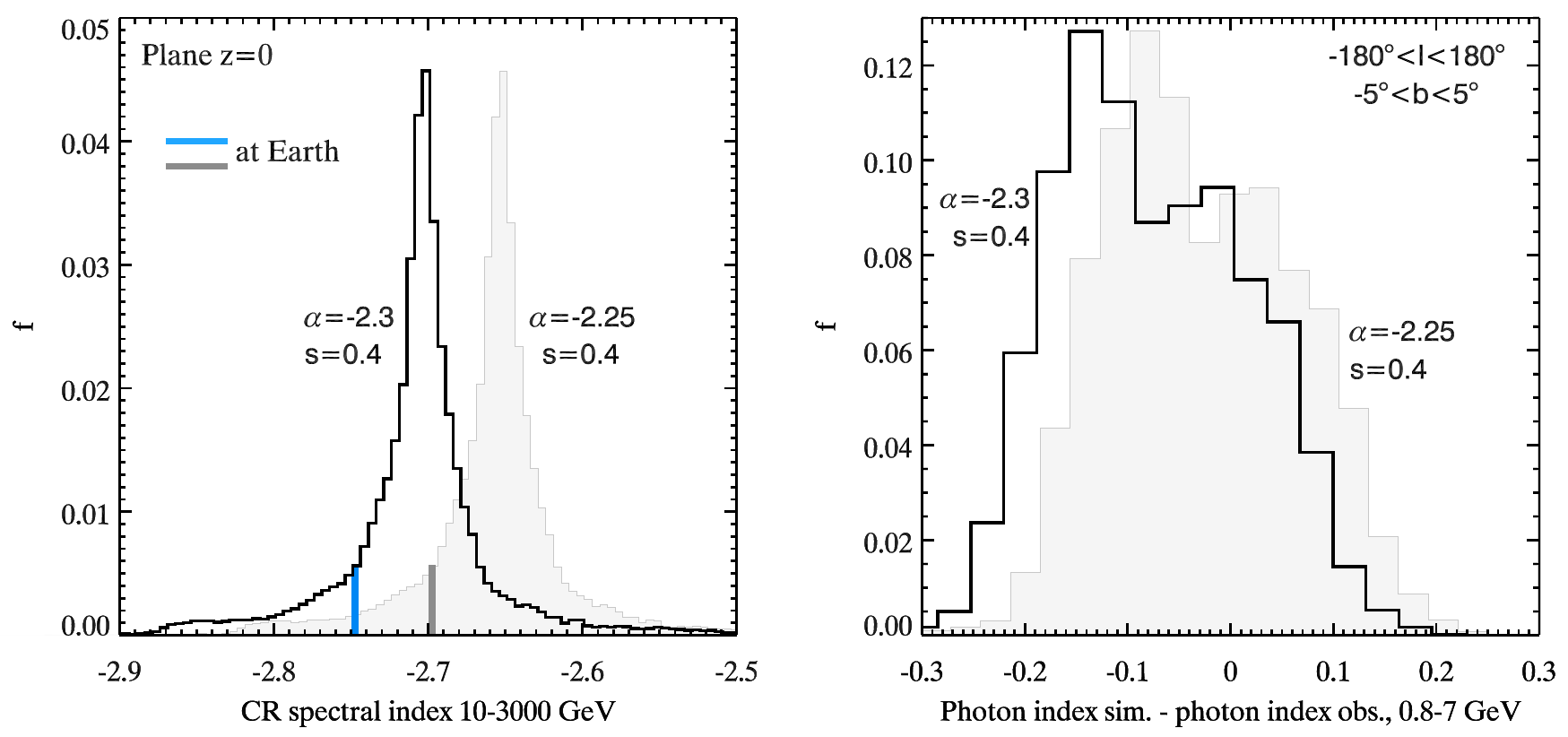}
\caption{Left-hand panel: the distribution of the simulated CR spectral index, estimated in the range $10-3000\,$GeV, in the plane $Z=0$. The open histogram shows the results assuming a injection spectral index $\alpha=-2.3$, while the shaded histogram is obtained for a harder injection index, $\alpha=-2.25$. In both cases, the diffusion coefficient is proportional to $E^{0.4}$. The vertical blue and grey lines depict the observed spectrum on the Earth, respectively. Right-hand panel: the distribution of the difference between the \gr\ spectral index in the plane of the sky obtained from simulations and from observations. A small latitude region has been selected: $-5^\circ<b<5^\circ$. The two different histograms respectively refer to the two different choices for CR injection spectral index. }
\label{fig:histo}
\end{figure*}

Fig.~\ref{fig:histo} (left-hand panel) shows the histogram of the CR spectral index values in the plane $Z=0$, estimated in the energy range $10\,$GeV-$3\,$TeV. As expected, the distribution is peaked at a value $\sim-2.7$ (open histogram). We note however, that the slope at the location of the Earth is somewhat steeper ($\sim-2.75$), due to the fact that, in this simulation, the Earth lies closer to the downstream side of the arm. Namely, the local CR spectrum is not representative of the average CR Galactic spectrum for this choice of parameters. A slightly different location of the Earth (or, a slightly different choice of the parameters describing the dynamics and/or morphology of the Galaxy) would allow us to recover a value equal to $-2.7$. However, we note that the results for \gr\ spectra present a similar feature---they are too soft as compared to observations. We compute the \gr\ spectral index in the energy range $0.8-7\,$GeV at different Galactic coordinates ($l$, $b$) and, for each position in the sky, we estimated the difference between the spectral slope of simulated and observed \gr\ spectra  (two examples of observed and simulated \gr\ spectra can be found in Fig.~\ref{fig:comparison_spectra}, dashed and solid lines, respectively). 
For the observed spectra, the slope has been computed in the same energy range, from \cite{Selig2015}. The results are shown in Fig.~\ref{fig:histo}, right-hand panel. The open histogram shows the distribution of the difference between the simulated and observed slopes. We limit this investigation to the region of the map at small latitudes, $-5^\circ<b<5^\circ$.
The average value is around $-0.08$, indicating that the simulated \gr\ spectra are on average softer than what inferred from observations.
To solve the inconsistency between observations and both the simulated CR spectrum at Earth and the \gr\ spectra, we consider a model with a harder injection spectrum $\alpha$.
The shaded histograms in Fig.~\ref{fig:histo} show that in this case predictions of the CR slope around the location of the Earth (left-hand panel) and of the map of \gr\ radiation (right-hand panel) are in better agreement with observational constraints.
For this reason, from now on we present and discuss in detail only the results obtained with $\alpha=-2.25$.
Fig.~\ref{fig:histo} (left-hand panel) also shows that the predicted CR spectrum at the location of the Earth (thick vertical line) is not representative of the average CR spectral slope (peak of the distribution). 
While in these examples the difference between the local and average spectral slopes is only 0.05, larger differences can be obtained by slightly changing the location of the arm as compared to the Earth. In this sense, the model is not predictive, and there is degeneracy among the simulation parameters. Although a complete study of the parameter space is beyond the scope of this paper, we outline that the proposed model gives a natural explanation for the increasing evidence that the local CR spectrum might not be representative of the average Galactic one \citep{ackermann12,neronov15}.

\begin{figure*}
\includegraphics[scale=0.8]{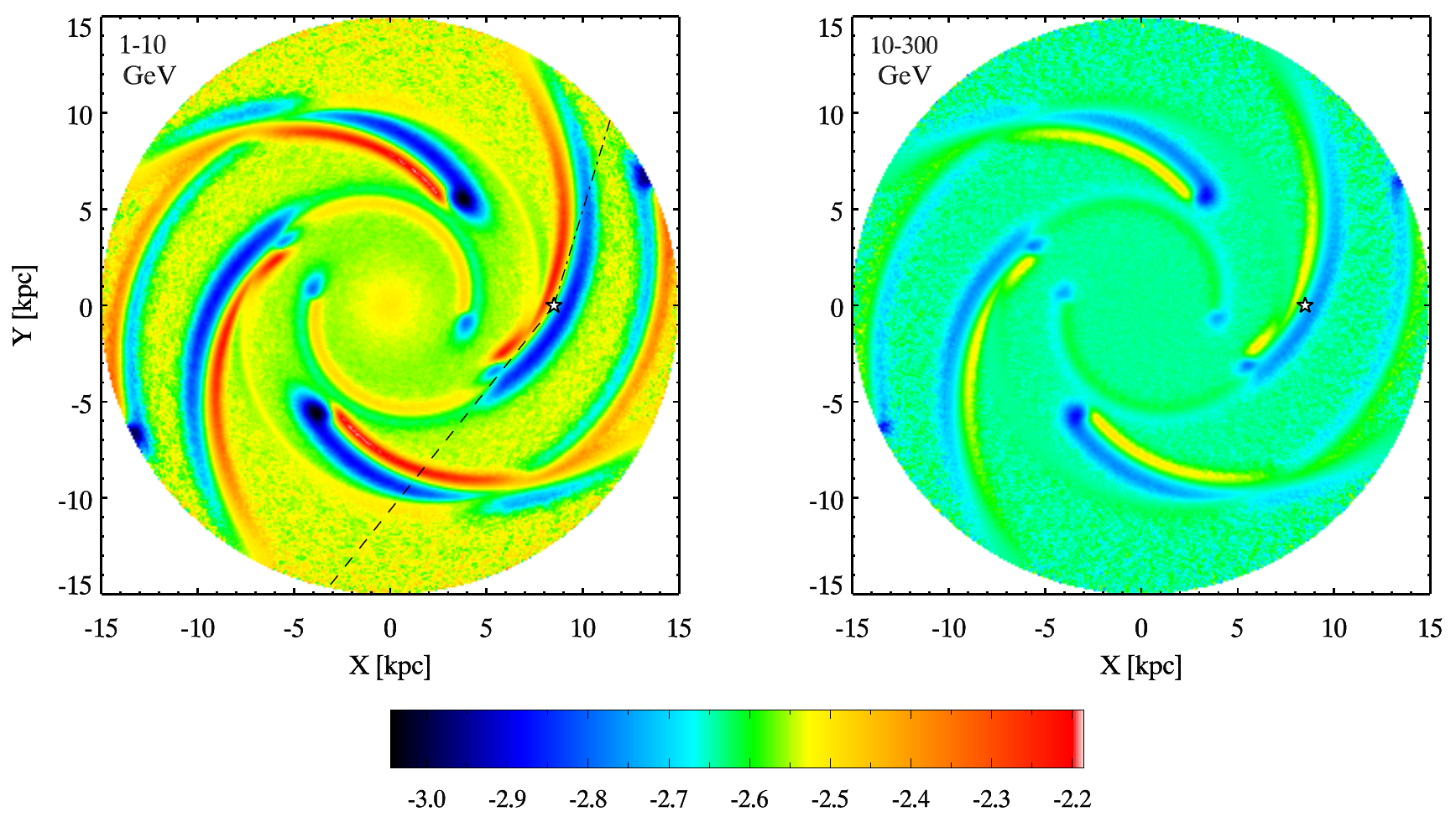}
\caption{A top view map of the Milky Way (at $Z=0$) showing the photon index of the \gr\ spectrum from $\pi^0$ decay, as measured in the range $1-10\,$GeV (left-hand panel) and in the range $10-300\,$GeV (right-hand panel), as obtained in the small halo model. The Sun is located at $X=8.5\,$kpc and $Y=0$ (star symbol). The dashed and dot-dashed lines show the two lines of sight that intercept the spiral arm.}
\label{fig:ph_slope_xy}
\end{figure*}

The spatial distribution of \gr\ spectra is expected to follow a pattern similar to the one characterizing proton spectra. 
As a consequence, also the \gr\ emission integrated along different \los\ will show variations in its spectral shape, depending whether the \los\ intersects harder or softer regions.
In order to quantify how the spectrum of the \gr\ radiation from \po\ decay changes across a spiral arm, we infer the spectral index $\delta$ by modelling the simulated photon spectra with a power-law function: $dN/dE\propto E^{\delta}$.
Since we expect a different behaviour at low and high energies, the spectral index is estimated in two different energy ranges: 1-10$\,$GeV and 10-300$\,$GeV. We begin by analyzing the small halo model (with a halo size that reproduces the Boron to Carbon ratio in the dynamic spiral arm model). 

Fig.~\ref{fig:ph_slope_xy} shows the map of the photon spectral index $\delta$ in the plane of the disc ($Z=0$) in the low and high energy ranges (left and right-hand panel, respectively).
The Sun is located at $(X,Y,Z)=(8.5\,\rm kpc,0,0)$, as indicated by the open star symbol.
We find that the spectrum is harder in the upstream side of the arms, and softer in the downstream side. This should be expected because lower energies are more strongly affected by the advection of the ISM through the arms. As a consequence, lower energy CRs are more abundant downstream, making the spectrum softer, and less abundant upstream, making the spectrum there harder.  Moreover, the variation of the spectral index across each arm is more evident in the lower energy range of 1-10$\,$GeV (where we find $-3.0<\delta< -2.2$) and becomes less evident at higher energies ($-2.9<\delta< -2.5$), where the relative effect of the advection, as compared to diffusion, is less important. 

Note that models including the spiral arms but neglecting their motion \citep{ShavivNewAstronomy,Pamela,DRAGONspiral} would instead predict the same spectrum across the Galaxy, with no arm signatures. 

\begin{figure}
\centerline{\includegraphics[width=8truecm]{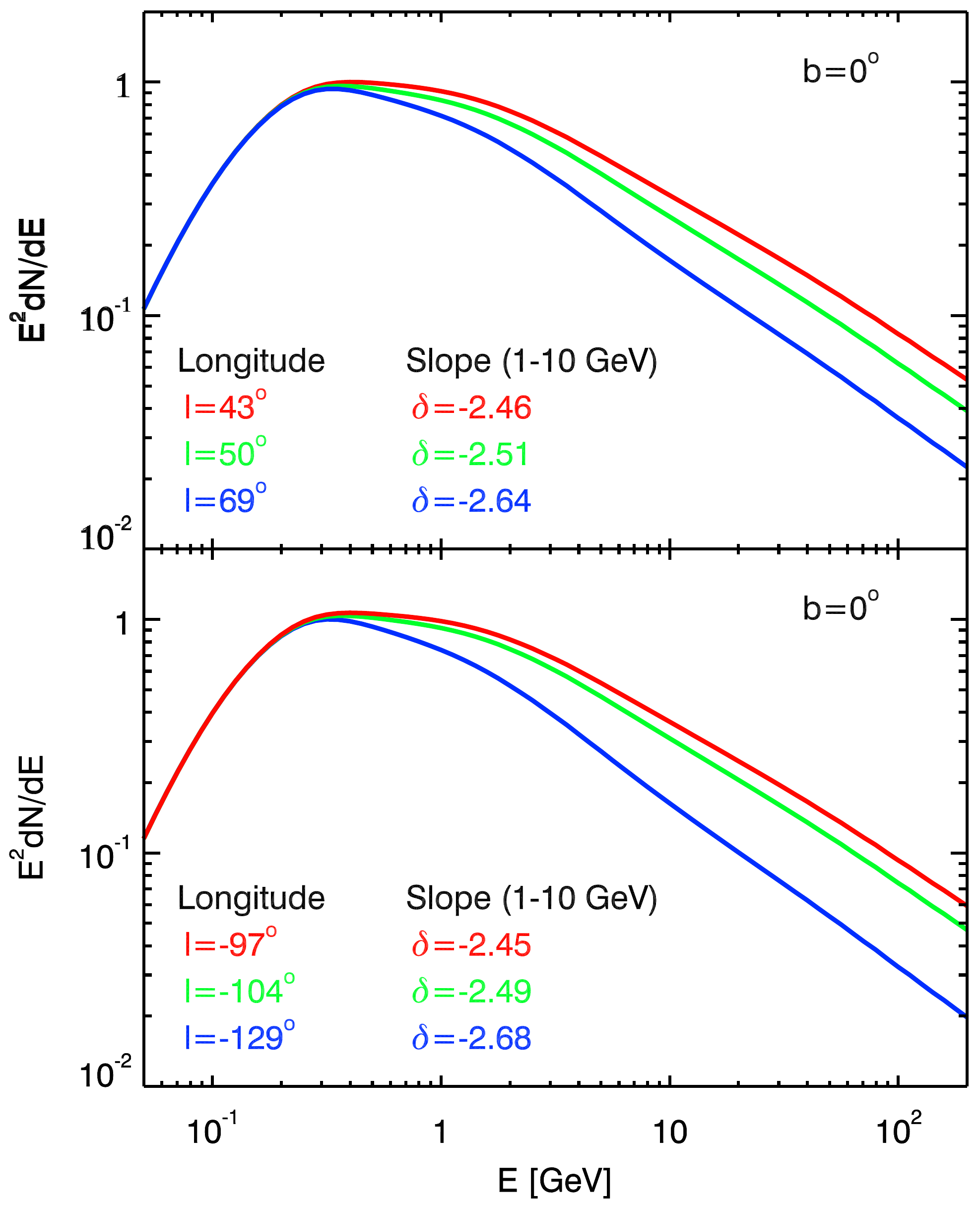}}
\caption{
Simulated $\gamma$-ray spectra at Earth from $\pi^0$ decay obtained after integration along the \los\, in the small halo model. In each panel, three different \los\ are shown: in the direction of the spiral arm (green), and on both sides of the spiral arm (red and blue), at latitudes $b=0^\circ$. The upper panel refers to directions around $l\sim50^\circ$ (dashed line in Fig.~\ref{fig:ph_slope_xy}), while the bottom panel refers to $l\sim-104^\circ$ (dot-dashed line in Fig~\ref{fig:ph_slope_xy}). For each spectrum, the longitude and the spectral index (in the range $1-10\,$GeV) are reported. Units on the $y$-axis are arbitrary, and the spectra have been normalized so that they overlap at low energies.
}
\label{fig:spectra}
\end{figure}
The variation of the spectral index across the arms affects the \gr\ spectrum detected at Earth. The effect is more pronounced near the directions of the arm tangents. 
Consider for example the closest arm and the arm tangents identified by the dashed line in Fig.~\ref{fig:ph_slope_xy} (left-hand panel). In the range 1-10$\,$GeV, the \gr\ spectrum measured at Earth is expected to have a spectral index around $\delta\sim-2.5$, as can be guessed from Fig.~\ref{fig:ph_slope_xy}. 
A slightly different \los, pointing now to either side of the arm would correspond instead to a softer or harder spectrum, depending on which side of the arm we are looking at.
The same effect is expected when the almost opposite direction is considered (i.e., the one represented by the dot-dashed line in Fig.~\ref{fig:ph_slope_xy}).

The spectra derived from integration along those \los\ that mostly intersect the centre of the arm are shown in the upper and lower panels of Fig.~\ref{fig:spectra} (green curves). The upper (lower) panel refers to the \los\ represented by the dashed (dot-dashed) line in Fig.~\ref{fig:ph_slope_xy}, and corresponds to a Galactic longitude $l\simeq50^\circ$ ($l\simeq-104^\circ$) and latitude $b=0^\circ$. 
Also depicted in Fig.~\ref{fig:spectra} are the spectra on both sides of the arm (red and blue curves). The longitudes and the spectral slopes (in the range 1-10$\,$GeV) are reported in each panel, while the latitude is always $b=0^\circ$ in all these examples.
Spectra are normalised so that they overlap at low energies. 

From the comparison between Figs.~\ref{fig:ph_slope_xy} and \ref{fig:spectra} it is evident that when integration along the \los\ is performed, the difference between the spectral slopes at the two different sides of the arm is strongly reduced. This is due to the fact that for $b=0^\circ$, the emission coming from more distant regions of the disc is also playing a role in shaping the spectra.

\begin{figure}
\includegraphics[width=\columnwidth]{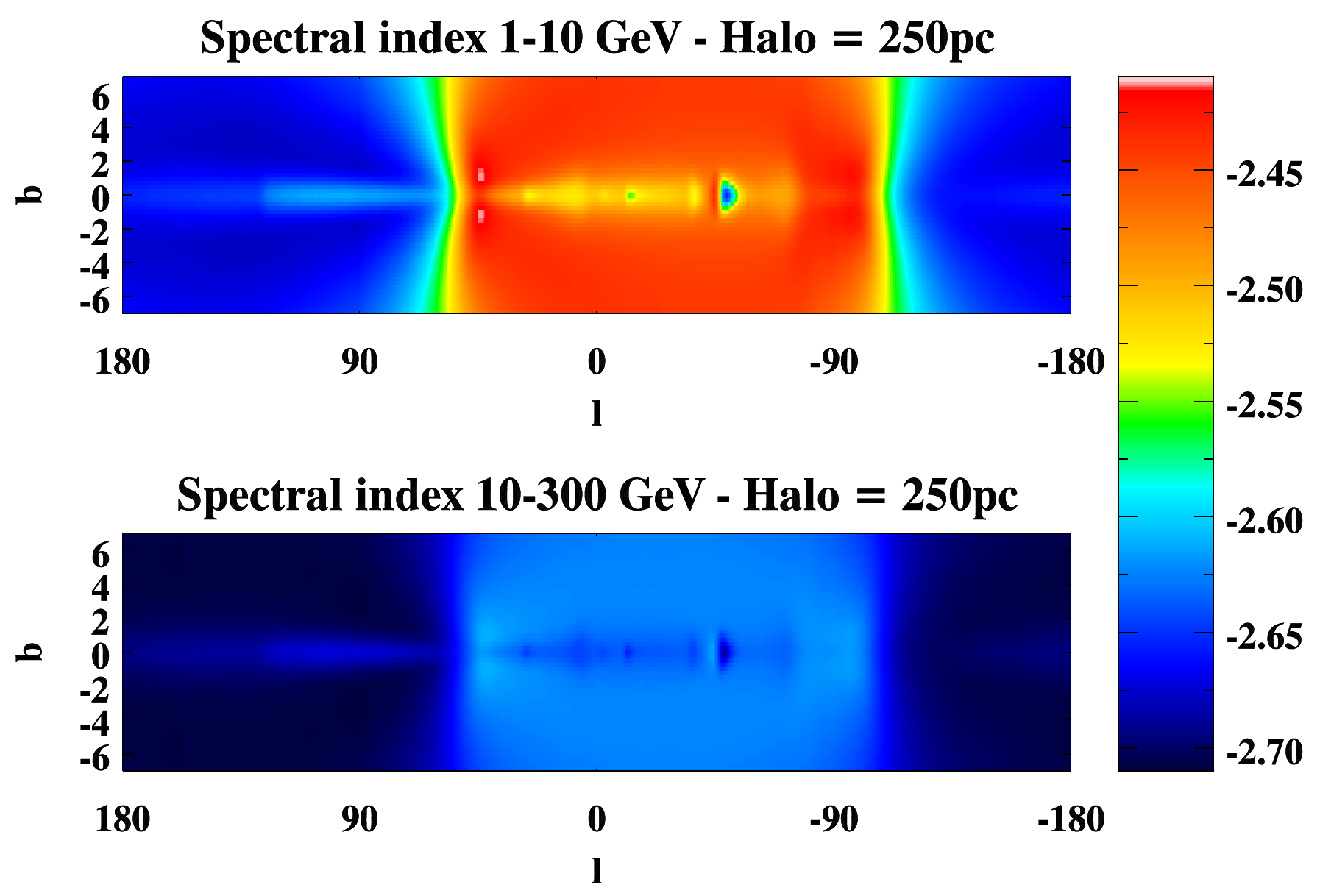}
\caption{Maps of the spectral slope in the energy range 1-10$\,$GeV (top) and 10-300$\,$GeV (bottom) of the $\pi^0$-produced $\gamma$-ray radiation, in the small halo model. The contour levels are denoted by the colour bar on the right, and it is the same for both maps.}
\label{fig:spectral-slope-map}
\end{figure}

We now extend the study of the spectral index to all the different directions of the sky, and build the spectral maps in different energy ranges.
The results are shown in Fig.~\ref{fig:spectral-slope-map} in the range 1-10$\,$GeV (upper panel) and 10-300$\,$GeV (bottom panel), for latitudes $|b|<7^\circ$.
The general trend is similar in both energy ranges, although the variation in the spectral index is much smaller at the higher energies ($\Delta\delta\sim0.1$).
With the help of Fig.~\ref{fig:ph_slope_xy}, we can understand the origin of the large variation in the low energy map (upper panel, Fig.~\ref{fig:spectral-slope-map}) as a function of longitude and latitude as follows:
\begin{itemize}
\item $|b|<2^\circ$ corresponds to \los\ that cross the arm, but are dominated by the disc component. For this reason the spectral index has an intermediate value ranging between $-2.5$ and $-2.6$.
In particular, as can be understood from Fig.~\ref{fig:ph_slope_xy}, at longitudes $-100^\circ<l<50^\circ$, $\delta\sim-2.5$, because the hardest part of the arm is crossed, while at $l>50^\circ$ and $l<-100^\circ$ the softest part of the arm is crossed, and $\delta\sim-2.65$;\\

\item $|b|>2^\circ$ at these latitudes, contrary to the previous case, the intersection between the \los\ and the disc component is reduced, and the contribution from the arm is dominating the emission, irrespective of the longitude, and it determines the spectral shape. The variation as a function of the longitude is determined by which side of the arm is the \los\ crossing. This explains why the spectrum is harder at $-100^\circ<l<50^\circ$, with an index close to $\delta\sim-2.4$, while for larger values of $|l|$ the spectrum is softer ($\delta\lesssim-2.65$). \\
\end{itemize}

Moving towards progressively higher latitudes, the spectrum tends towards softer values, at all  longitudes. The regions characterized by the most extreme (soft and hard) spectral indices are in fact missed by the \los\ and an intermediate spectral index is recovered.

Summarizing, two main features are predicted by our small halo, dynamic spiral arm model. 
Spectra are hard at small/intermediate Galactic longitudes (corresponding to the central part of the map), since these longitudes correspond to \los\ crossing the side of the arm opposite to the direction of the motion (red in Fig.~\ref{fig:ph_slope_xy}). 
On both sides of this central hard region, where latitudes are higher, 
spectra are soft, since these latitudes correspond to \los\ that cross the other side of the arm (blue in Fig.~\ref{fig:ph_slope_xy}).
The second prediction is that this trend is particularly evident at latitudes $2^\circ<b<4^\circ$, where the contribution from the arm is dominating the emission, and is reduced both at very small ($b<2^\circ$) latitudes (where the emission is the result of contribution from both the arm and the disc) and at large latitudes ($b>4^\circ$), where the local emission governs the spectral slope.

\begin{figure}
\includegraphics[scale=0.465]{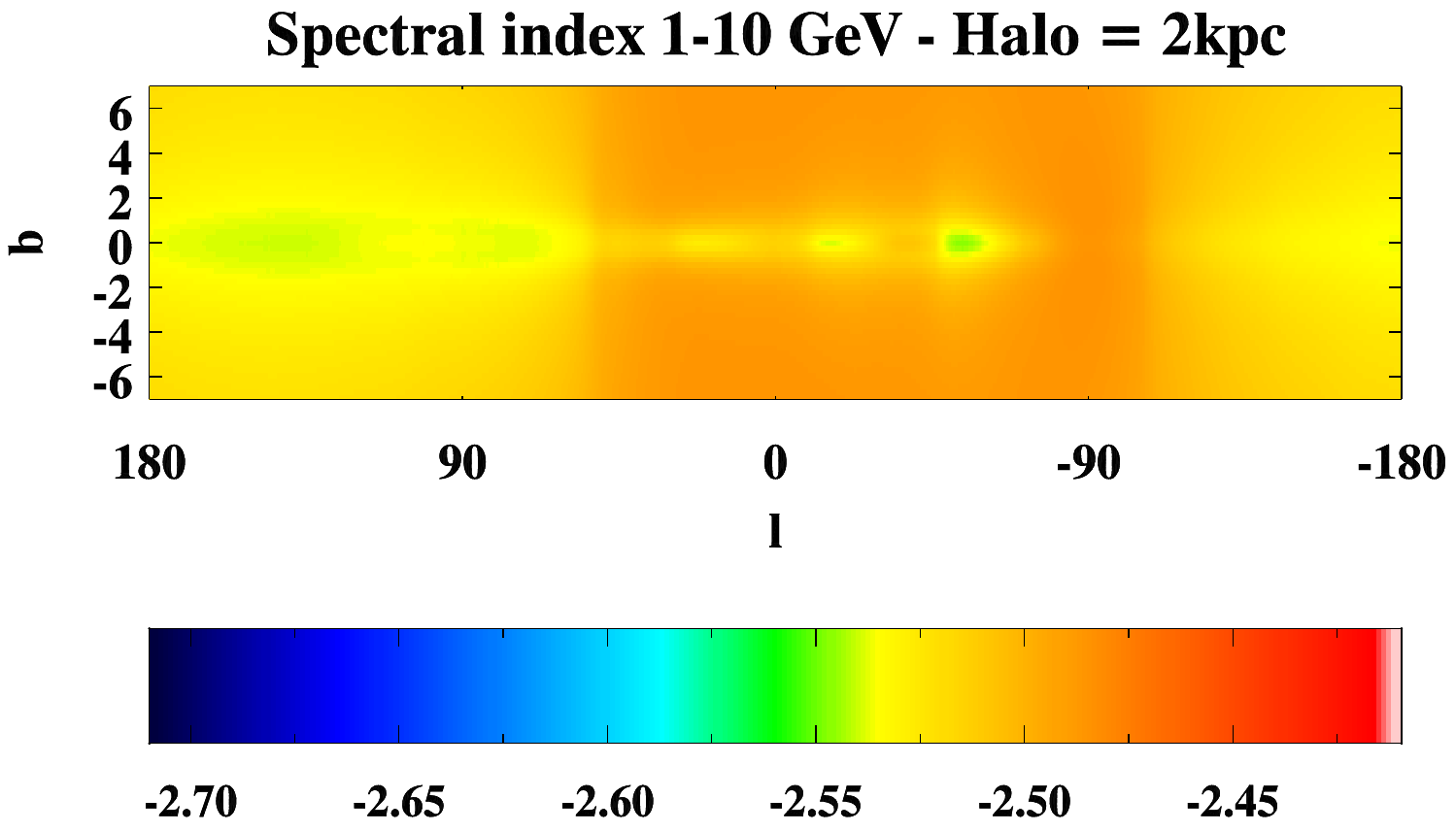}
\caption{Map of the spectral slope in the energy range 1-10$\,$GeV of the $\pi^0$-produced $\gamma$-ray radiation obtained from the large halo simulation ($Z_h = 2\,$kpc). The contour levels are denoted by the colour bar on the bottom, and it is the same as in Fig.~\ref{fig:spectral-slope-map}.}
\label{fig:large_halo}
\end{figure}

Fig.\ \ref{fig:large_halo} depicts the prediction of the large halo model, with $Z_h=2\,$kpc, in the energy range $1-10\,$GeV. When compared with the small halo model (Fig.~\ref{fig:spectral-slope-map}, upper panel), the same qualitative features exist, but quantitatively they are different. 
First, instead of a well defined ``oval" with a higher spectral index, the respective oval is significantly less defined. Second, instead of having variations in the index of order $\sim 0.2$, there are variations of only $\sim 0.05$. 

The differences between the two different halo size models are not surprising given that the large halo model requires a larger diffusion coefficient (in particular, to fit the ``age" derived from the Beryllium isotope ratios). As a consequence, characteristics of the CRs in the vicinity of the Galactic arms are smeared over large regions, wiping away details.

\section{A Comparison with observations}
\label{sec:observations}
\begin{figure}
\includegraphics[scale=0.42]{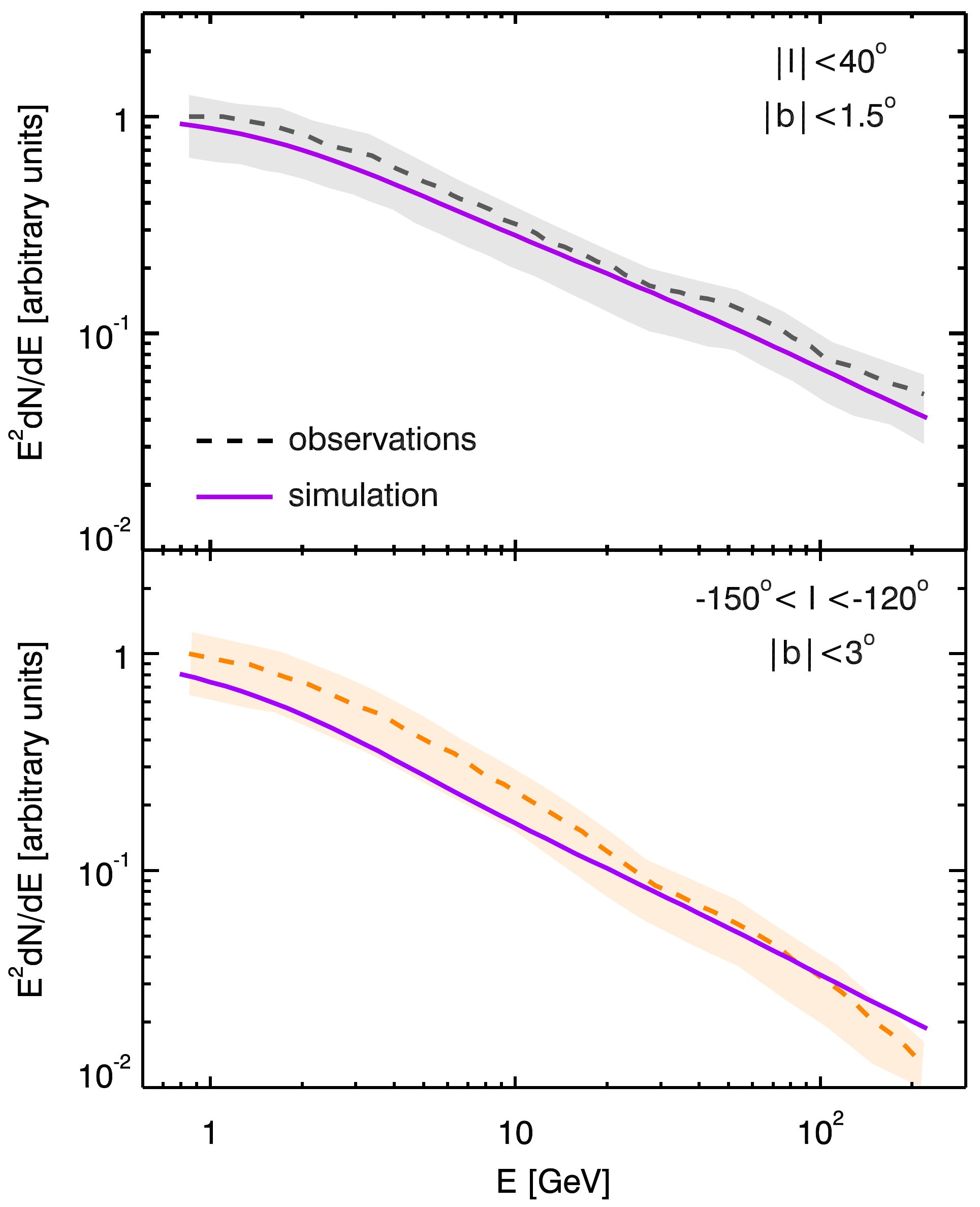}
\caption{Comparison between observed and simulated \gr\ spectra in two different regions of interest: $|l|<40^\circ$ and $|b|<1.5^\circ$ (upper panel), and $-150^\circ<l<-120^\circ$ and $|b|<3^\circ$ (lower panel). 
Observations (taken from \citealt{Selig2015}) are shown with a dashed line.
The shaded areas represent the uncertainties, including statistical and systematic errors. The spectrum derived from our simulation in the same region of the sky is instead shown with a solid line. Normalizations are arbitrary.}
\label{fig:comparison_spectra}
\end{figure}

Recently, a spectral map (as measured by {\it Fermi}) of the anisotropic component (dominated, especially at low energies, by $\pi^0$ decay) of the diffuse $\gamma$-ray emission has been presented by \cite{Selig2015}, allowing us to compare our numerical results with observations.
Their analysis is performed in nine energy bins, from 0.6 to 307.2~GeV.
First, we compare their measured \gr\ spectra estimated in two different regions of the sky with our simulations. The results are shown in Fig.~\ref{fig:comparison_spectra}: data from \cite{Selig2015} are shown as a dashed line. Uncertainties are represented with a shaded area. The spectrum derived from our simulation is instead shown with a solid line. The upper and bottom panels refer to two different regions of the sky. Normalizations and units on the $y$-axis are arbitrary. An inconsistency between the simulated spectrum and the observed one in the outer region is evident at energies $>10\,$GeV, where the data show a steepening of the spectral slope. This steepening found by \cite{Selig2015} is inconsistent with results from \cite{ackermann12} and the authors identify the origin of this inconstancy with the low signal to noise ratio in the highest energy bins.

We use the fluxes provided in the first four energy bins\footnote{https://www.mpa-garching.mpg.de/ift/fermi/} (corresponding to the energy range 0.85-6.79$\,$GeV) and estimate the spectral index by fitting the data with a simple power-law function, consistently with the method adopted to derive the slope of the simulated spectra. 

\begin{figure}
\includegraphics[width=\columnwidth]{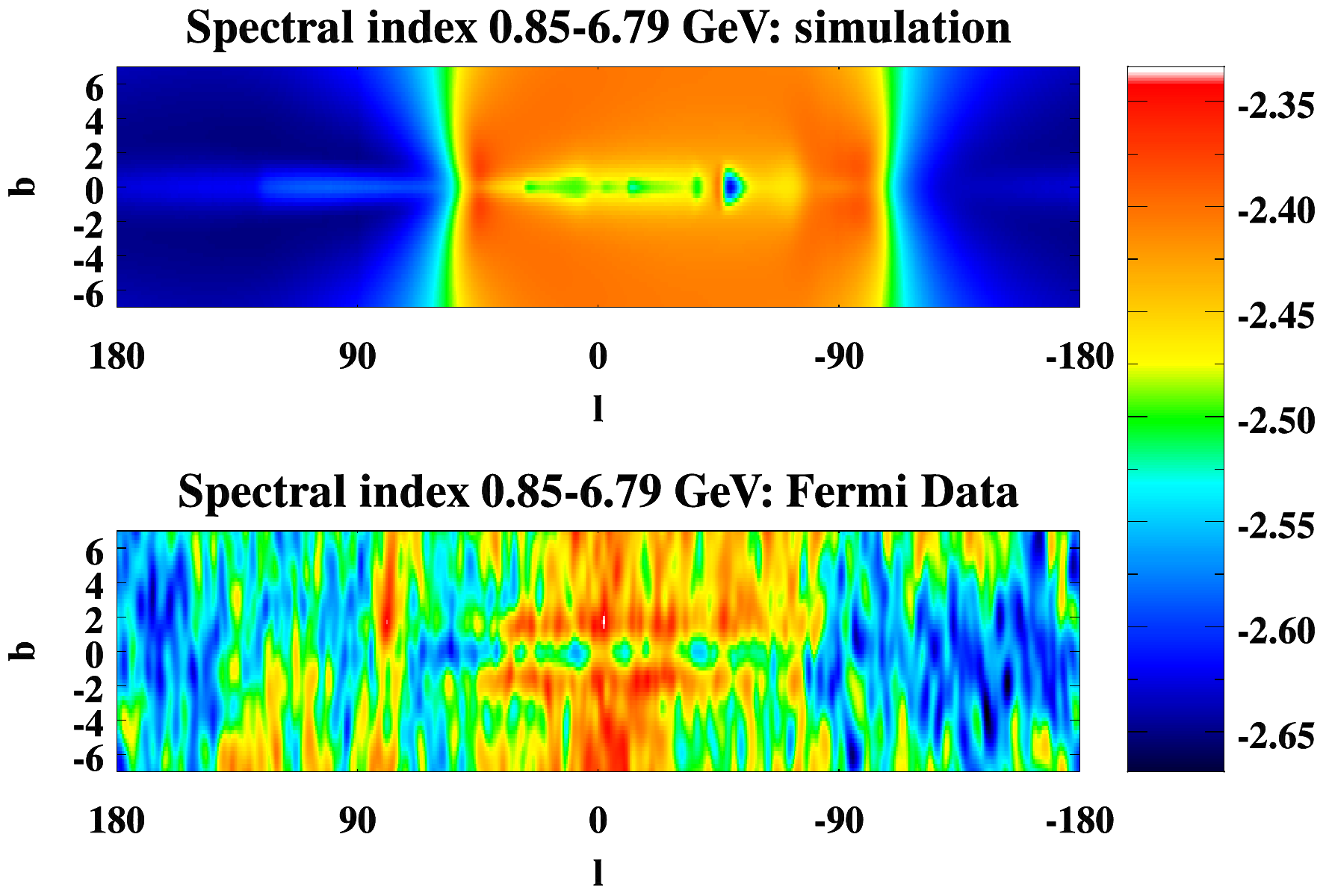}
\caption{Maps of the spectral slope of the $\pi^0$-produced $\gamma$-ray radiation in the energy range 0.85-6.79$\,$GeV. 
The upper panel depicts the results from  the small halo  simulation, while the bottom panel is the {\it Fermi} map adapted from \citet{Selig2015}. 
The same color bar is used for both the simulated and the observed maps.
}
\label{fig:comparison_wdata}
\end{figure}

The {\it Fermi} spectral map at $|b|<7^\circ$ in the range 0.85-6.79$\,$GeV is depicted in the lower panel of Fig.~\ref{fig:comparison_wdata}. 
We focus on this range of energies and latitudes because it is where the diffuse emission of Galactic origin is expected to be dominated by $\pi^0$ decay, and the comparison between our simulations and observations is therefore justified. 
In order to perform a proper comparison, we estimate the spectral index of our simulated spectra in the same energy range.
The simulated map for the small halo model is showed in the upper panel of Fig.~\ref{fig:comparison_wdata}.

First of all, we note that the range of $\delta$ values in the simulated and observed spectra is similar: $-2.65<\delta<-2.35$.
Moreover, the two maps are very similar, i.e., $\delta$ varies as a functions of $b$ and $l$ according to a similar pattern.
All the features present in the simulated map and discussed in the previous section are visible also in the map derived from {\it Fermi} data, albeit with some measurement noise. 
This allows us not only to assess the validity of our numerical results, but also to provide an interpretation for the observed spectral map presented by \citet{Selig2015}, at least in this range of latitudes.

By inspecting the {\it Fermi} map in the region $-90^\circ<|l|<40^\circ$, we note that the spectra are on average harder than the spectra on both sides of the central region. Moreover, in this central part of the map, a variation of the spectral index with the latitude is visible.

At $|b|<1^\circ$ the spectral index has an intermediate value, and it becomes harder at around $|b|\approx2^\circ-3^\circ$. We have interpreted this variation of $\delta$ as due to the transition from \los\ where the contribution from the disc is relevant, to \los\ where the closest arm is dominating the emission. At even higher latitudes the spectrum becomes softer, consistently with the fact that the contribution from the side of the arm becomes negligible again. 

Also the region at larger longitudes $|l|>40^\circ$ exhibits average properties very similar to those found in the simulated spectra. The spectrum is significantly softer, which we interpreted as due to the fact that these \los\ mainly intersect the downstream side of the arm.

While the small halo simulation reproduces the main features in the {\it Fermi} index maps, this is not the case with the large halo simulation, the results of which are depicted in Fig.\ \ref{fig:large_halo}. Specifically, the large halo model has significantly less defined features compared with both the small halo simulation and the actual {\it Fermi} observations. In other words, the latter strongly point towards a small halo. 

We have also compared the observations with the spectral map derived with $\alpha=-2.3$. In this case, a pattern similar to the one shown in Fig.~\ref{fig:histo} is recovered, but (as anticipated in \S\ref{sect:results}) the values are shifted towards softer spectral indices, and do not match the observed range of values observed by {\it Fermi}.

\section{Discussion and Summary}
\label{sec:discussion}

A spiral arm source distribution was first incorporated into a CR diffusion model to explain the apparent time variation in the CR history as reconstructed with Iron Meteorites \citep{ShavivNewAstronomy}. As a ``side effect" it was found to explain the Pamela anomaly in which the positron fraction appears to increase above about 10 GeV \citep{Pamela,DRAGONspiral}. These models, however, do not introduce any additional time-scale for the proton population, because of which the $\pi^0$ produced $\gamma$-rays should exhibit the same power-law index in every direction, unless new physics is introduced. 

Such new physics was introduced by \cite{Benyamin2014} who realized that if one considers the {\it dynamics} of the spiral arms, namely, that the ISM moves relative to the arms, the CRs undergo advection relative to the CR sources (in addition to diffusion). This advection naturally predicted a rise in the Boron to Carbon ratio at energies below 1 GeV/nuc., as observed. However, several additional consequences ensued from this analysis. First, because of the different path length distribution, it was found that in order to correctly recover the overall amount of secondaries produced, one must take a smaller halo and a smaller diffusion coefficient. Second, it was found to consistently explain the sub-Iron to Iron isotopes, which in standard models require an inconsistently different average column density \citep{Benyamin2016}. 

The next natural step was to simulate the $\gamma$-ray production through $\pi^0$ decay, and compare with observations. This is more robust than other $\gamma$-ray components that require the distribution of electrons and the magnetic field as well, and therefore left for future analysis.  In particular, given that standard models predict a fixed power-law index,  by studying directional variations of the power-law index we can concentrate on differences from the standard model. 

In the analysis, we have found that the dynamic spiral arm model exhibits interesting characteristics. First, there is a very clear upstream/downstream asymmetry, such that the upstream side of spiral arms is harder. In addition, the Galactic plane region inwards (towards the Galactic centre) of the two opposite  arm tangents is harder than the plane outside the tangents. Note that because the arm tangents are not symmetric around the Galactic centre, this asymmetry is not expected to be symmetric either. The {\it Fermi} data exhibit this signature--the spectral slope at the plane towards the centre of the Galaxy is harder than in the other directions. Moreover, the data appear to exhibit an asymmetry about the Galactic centre expected from the asymmetric arm tangents, though not unequivocally given the data quality. 

The second interesting signature, apparent in both the small halo simulation and the {\it Fermi} observations, is the vertical dependence. When observing between the arm tangents then the spectral index increases from the plane up to an altitude of about 1.5$^\circ$, and then decreases further away from the plane. This seemingly odd behaviour is due to the fact that at the Galactic plane, the \los\ crosses the harder upstream side of the arm but the significant contribution from the Galactic disc further inside the Galaxy flattens the spectrum. On the other hand, at a somewhat higher angle, the Galactic disc is mostly absent since the \los\ passes above it, and the contribution from the hard arm dominates the spectral shape. At even higher latitudes, the \los\ already passes above the arm and the local softer contribution is left to dominate. Since the aforementioned features are significantly attenuated in the large halo model that requires a larger diffusion coefficient, we can conclude that the {\it Fermi} index maps are surprisingly consistent with the small halo simulation, but not the large halo. 

It should be emphasized that unlike other models that have to add additional ad hoc assumptions to obtain spectral changes (e.g., on the diffusion characteristics inside and outside arms, or a contribution from dark matter), the dynamics of the arms is an unavoidable component that should necessarily be included. We know that the CR sources are in the arms and that the ISM is advected through them. 

Another interesting point is that because this model predicts that the CR spectrum varies across the Galaxy, the local CR spectrum measured at the Earth and the average CR spectrum inferred from the average of the \gr\ spectra are not necessarily the same. This is consistent with (and provides an explanation for) the apparent inconsistency between the local and average spectral slopes. 
The difference we need to introduce in our model is of 0.05, but larger differences can also be explained by slightly changing the location of the arm with respect to the Earth in the simulation.

One should also mention several caveats in the model.  First, the ISM density is the same in the spiral arms and outside of them. This is a naive approximation, the effect of which is not supposed to change the slopes in the XY map, however, it will emphasize the $\gamma$-rays emission coming from the arms, and therefore should somewhat increase the predicted spectral slope contrasts. On the other hand, the model does not consider possible variations of the diffusion coefficient (e.g., because of the magnetic field and ISM turbulence characteristics) that could give different spectral slopes in the arms and outside. 

Another interesting point to consider is the effect of the {\it Fermi} bubbles \citep{Su2010}. It is clear that at a high enough Galactic latitude it will dominate the spectrum \citep{Selig2015}. This was not considered here, but is should start dominating the spectral index at some latitude above the plane. 

Nevertheless, even with the caveats considered, the general agreement between the theoretical predictions and observations suggests that the CRs sources are indeed concentrated towards the arms (as expected from SNR), and that the halo size and diffusion coefficient are smaller than the canonical values arising in standard azimuthally symmetric models.  

\section*{Acknowledgements}
LN was partially supported by a Marie Curie Intra-European Fellowship of the European Community's 7th Framework Programme (PIEF-GA-2013-627715). TP was partially supported by Adv ERC grant TREX. 
NJS gratefully acknowledges the support of the IBM Einstein Fellowship of the Institute for Advanced Study, and Israel Science Foundation (grant no.\ 1423/15).  This research was also supported by the I-CORE Program of the Planning and Budgeting Committee and the Israel Science Foundation (center 1829/12). 
The authors acknowledge the use of the open access material presented by Selig et al. (2015) on the application of the D$^3$PO inference algorithm to the $\gamma$-ray sky seen by the {\it Fermi}-LAT.


\bibliographystyle{mnras}
\bibliography{Gamma-Ray-MWDisk-Refs}


\bsp	
\label{lastpage}
\end{document}